# Tile Multiple-Readout Compensated Calorimetry

David R Winn* -  Fairfield University
Yasar Onel -  University of Iowa

*Introduction*

We propose extending parallel fiber dual readout calorimetry to tiles, more applicable to many future experimental requirements, with superior energy resolution. Monte Carlo (MC) studies indicate that a tile dual calorimeter including an integral Cerenkov-compensated e-m front end and further longitudinal segmentation, not possible with parallel fibers, has equivalent or better resolution. Besides comparison and tuning with dual tile calorimeter data, a MC can be extended to study other dual and then multiple tile sensors including tiles with higher contrast to em-hadron shower fluctuations with low refractive indices (much lower than quartz or plastic), transition radiation, secondary emission, hydrogenous/non-hydrogenous ionization-sensing, and neutron and ion-fragment sensing tiles for improving dual readout not available with fibers, and beyond to triple or more readout. For example, secondary emission tiles (like dynodes) are very sensitive to ion fragments and low energy neutrons. We suggest MC studies for adding Cerenkov and other tiles to Particle Flow/High Granularity tile calorimeters such CALICE and planned in CMS & ATLAS Phase III upgrades, groups studying future machine(ee,pp,ep) detectors, b-physics, tagged neutrino experiments, and space-based calorimeters.

Calorimetric technologies are foreseen as essential in future frontier experiments, and require better hadron/jet energy and angle resolution, better isolation of electrons/photons (H-› γγ; b, t, W, Z-› leptons), better performance under pileup and over a wider range of η. Since $m_H$ is relatively light, at higher √s in future machines, hermetic η reach must expand. Even at LHC ~25% of H-› μμ and 90% of [H-› ZZ-› 4 leptons] have ≥1 lepton at η>2.8, requiring ID and energy measurement. Hermetic calorimeters are needed for missing $E_T$ measurements as low as feasible, consistent with neutrino escape and forward muon ID(total dE/dx). Vector boson scattering (WW, ZZ, WZ) where color is not exchanged is an important SM confirmation or BSM signal, requiring resolved forward tagging jets.  These signals will be important for S/LHC and beyond (70-100 TeV), and e⁻-e+ machines producing (H+Z), tt, and beyond.

Calorimeters using scintillator or Cerenkov light are standard techniques in experimental particle physics, including in precision, high intensity, high energy and particle astrophysics experiments. Many optical calorimetry techniques offer reasonably acceptable time resolution, rate capability, and energy resolution yet need to be improved. A particularly important example is jet energy resolution sufficient for identifying and distinguishing W-›jet-jet from Z-›jet-jet decays which requires $\sigma_E/E \leq 20\%/\sqrt{E}$. In intensity and energy frontier experiments, many scintillators and Cerenkov media could be useful except for radiation damage. Conversely, detector time resolution and rate capability are also hallmarks needed in future experiments, which some rad-hard calorimeter technologies (noble liquids, some slower inorganic scintillators) do not have. Future experiments and colliders require both sensors and embedded readout electronics radiation resistant to 10's or 100's of MRad and operation at ~100MHz and ~200 pileup.

High Granularity Particle/Energy Flow Calorimetry[1,2] and Dual Readout[3,4] are developing calorimeter technologies that when combined may be capable of unprecedented energy resolution, segmentation against pileup, excellent time resolution and high rate using ionization sensor tiles and Cerenkov tiles, provided that tiles of sufficient speed and radiation damage resistance can be employed, with a readout method also rad-hard. Dual Readout is a technique to correct calorimeter response functions by measuring as separately and independently as possible the fluctuating electromagnetic and hadronic components of a hadron or jet calorimeter shower, event-by-event, using the 2 pieces of information to correct the energy, leading to a narrower and more Gaussian energy spectra.

* Corresponding author – winn@fairfield.edu ORCID 0000-0003-2637-5743

At present, Dual Readout for jet calorimeters has only been explored using parallel plastic scintillator fibers and quartz Cerenkov fibers, and has achieved $\sigma_E/E \sim 30\text{-}35\%/\sqrt{E}$, when pitched at a few degrees to the beam. Preliminary MC studies by our group indicate hadronic energy resolution $\sigma_E/E \sim 18\%/\sqrt{E}$ is possible from Tile Dual Readout. In contrast to parallel fibers, tile readout can be designed with sufficient granularity to enable particle/energy flow and could have pileup capability and high radiation damage resistance unavailable with existing parallel fiber dual readout. We therefore propose to explore and optimize a straight forward realization of Dual Readout using scintillator tiles and Cerenkov tiles instead of parallel fibers, capable of approaching the predicted $18\%/\sqrt{E}$ AND capable of being extended to high granularity for the challenging environment of future colliders. We also propose studying the potential for even better hadronic compensated energy resolution by adding multiple tile readouts – 3rd or 4th sensor with different responses to shower properties (e-m, hadron, neutrons, ion fragments).

*Frontier Applications of Precision High Rate Radiation Resistant Calorimeters:*
*Energy Frontier:*
- *Identification and Separation of W,Z→jet-jet decays:* Future experiments would benefit from reconstructing/identifying Z's & W's by jet-jet decays, 5-6 times more numerous than leptonic decays *and especially to separate W→jet-jet decays from Z→jet-jet decays*. Reasonable separation requires a relative jet energy resolution of ~3% at 100 GeV, with typical jet single particle energies ~10-15 GeV. A jet energy resolution of ~3% over 50-500 GeV yields a 2.6-2.3$\sigma$ W/Z separation (Fig 1)[5].

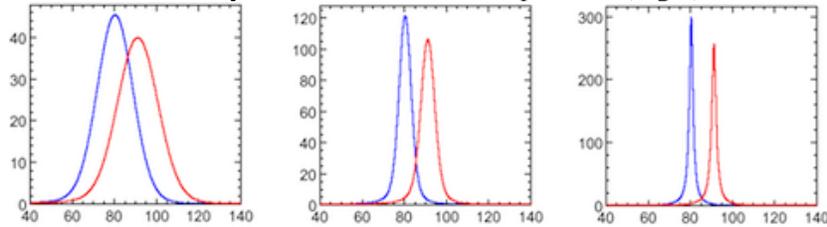

*Fig. 1a,b,c*: W(blue)/Z(red)→jet-jet mass separation: *left* - typical hadron calorimeter $\sigma_E/E=60\%/\sqrt{E}$; *middle* $\sigma_E/E = 3\%$ at 50 GeV with 2.6$\sigma$ separation; and *right* -perfect resolution with ~4.5$\sigma$ separation.

- *Vector Boson Scattering and Fusion:* WW,WZ, ZZ – jet-jet decays and including forward tagging jets, when color is not exchanged (pp→VV+tag jet+tag jet+X), with high radiation resistance.
- *Missing $E_T$: Dark-matter, SuSY-* $\sigma E_{Tmiss}$ as low as feasible, consistent with neutrino escape, requiring best possible energy-resolution+particle flow (→100*MHz collision rates, pileup >100/crossing*)
- *Exotics:* Z'/W'→ jet-jet
- *Calorimeter Issues:* As discussed below, a single jet calorimeter tower with parallel fibers will have raddam in the front section of the calorimeter; a longitudinally segmented tile calorimeter is amenable to replacing damaged front sections, and to have a multiple readout compensated e-m front end.
– **Intensity Frontier**: requiring excellent calorimeter energy and angle resolution in future LHCb, $e^+e^-$ b-factories, beam-dump dark photon/DM, and high energy neutrino scattering at future μ/ν factories.
- *b, τ via jet-jet decays* in b-factories, b-physics pp(LHCb) or ep collider experiments.
- *τ detection via jet-jet decays*- Long Baseline(LB) high energy ν Detectors (muon factory $\nu_\mu$ beams).
- *LB ν/Atmospheric ν:* Combined Cerenkov light + ionization (drifted LAr) in large detectors
- *Dark Photon/DM experiments* -  Ex: 1-10 GeV tagged electron beams to detect small missing $p_T$.
- *Tagged $\nu_e$ Beams from K-decay* –requires high rate (>100 MHz) and rad-resistant calorimetry[6,7]
- **Cosmic Frontier:** - Future balloon or AMS-like space-based (as proposed for the China Space Station[8]) experiments: more compact calorimeters yet with similar resolution, when combined with particle-flow.

*Dual Readout – Precursor to Multiple Readout*
Dual Readout is a technique to correct calorimeter response functions by measuring separately the fluctuating electromagnetic and the hadronic components of a hadron or jet calorimeter shower, event-by-event, and using the 2 pieces of information to correct the energy, leading to a narrower and more Gaussian energy spectra. This technique is a precursor to multiple readout with particle flow/high granularity readout that we propose here. One method, Cerenkov Compensation, is a Dual Readout technique that approximates separate measurements of e-m and hadronic components by simultaneously measuring Cerenkov light(e-m) and ionization scintillation(hadronic) light, but only *quasi*-independently in calorimeter showers. This compensation technique was invented and studied in detail using a GEANT MC in 1988 by our group ["Compensating Hadron Calorimeters with Cerenkov Light", D. R. Winn and W. Worstell, IEEE Trans. Nuclear Science Vol. NS-36 , No. 1, 334 (1989)]. In that MC study, drifted ions and Cerenkov were simulated in LArgon. Cerenkov compensated Dual Readout technique has been reduced to practice by R.Wigmans et al. from Texas Tech (TTU)[9]. The TTU group used parallel fiber geometry, with parallel, longitudinal quartz Cerenkov fibers and plastic scintillating fibers in a Cu matrix, based in part on our group's history of development of parallel scintillating fiber and Cerenkov fiber calorimetry for SSC, and used as a rad-hard Cerenkov fiber forward calorimeter in CMS. It achieved a hadron energy resolution of $\sigma_E/E \sim 30\%\text{-}35\%/\sqrt{E}$, worse than the potential resolution of Dual Readout, as we will discuss in following sections.

*Rules of Thumb for Multiple Readout Calorimeter Resolution*:

(0) *Limit:* An intrinsic limit of sampling hadron calorimetry is $\sigma_E/E > 11\%\text{-}13\%/\sqrt{E}$, given by the ratio of detectable neutron energy to the fluctuations lost in nuclear binding energy.

(1) *Contrasts:* The contrast between hadronic energy h- and e-m energy e- energy signals should be as large as feasible. Typically these are parametrized by $h/e_i$ (i=ionization) and $h/e_C$ (C=Cerenkov). The ratio of ratios $[h/e_i]/[h/e_C]$ must be $\geq 4$ in order to reach incident hadron energy resolutions well-below $30\%/\sqrt{E}$, with $18\%/\sqrt{E}$ being a reasonable target to achieve using plastic scintillator and low index Cerenkov materials[10];

(2) *Hadronic signal:* $h/e_i$ as large as possible - implying hydrogenous and/or neutron-sensitized ionization detection media;

(3) *E-M resolution:* The intrinsic resolution on e-m energy from Cerenkov light must be $< 80\%/\sqrt{E}$ to achieve jet energy resolution $< 20\%/\sqrt{E}$;

(4) *Sampling Energy Resolution:* In general, energy resolution scales $\sim \sqrt{(f_{sample}/f_{frequency})}$.

(5) *Compensation* can also be approached by: 1. enhancing the neutron sensitivity (hydrogenous media, or fissionable media such as Uranium[11]) and/or 2. suppression of electromagnetic components by tuning the absorber thickness relative to sampling media ($f_{sample}$ typically ~1/10), but at a loss of potential ultimate resolution. An example: compensated ($E_{em} \sim E_{had}$) Cu+Scintillator has $\sigma_E/E \sim 60\%\sqrt{E}$.

(6) *Multiple Sensors:* Adding sensor tiles which are relatively insensitive to MIPs from e-m showers and more sensitive to $\gamma\beta \to 0$ such as ion fragments will increase the contrast between e-m and hadronic energy by enhancing the low energy hadronic signal. An example is Secondary Emission; its signal scales as dE/dx, with a MIP SE signal ~100x less than that of the energy of the peak signal. The peak signal for protons($\pi\pm$) occurs at ~200(20)KeV. Neutron enhancement can occur by n+p->p+n knock-on protons, spallation and fission. Similarly, lower index n tiles or TRD have less sensitivity to hadrons and neutrons.

(7) *Non-hydrogenous dense inorganic scintillators* (LYSO, $PbWO_4$, $CeF_3$, etc) have $h/e_i \sim 0.4$ and $h/e_C \sim 0.25$, or $[h/e_i]/[h/e_C] \sim 1.6$, and as such, homogeneous calorimeters (without inert absorber) measuring simultaneously scintillation and Cerenkov light *cannot* achieve dual readout compensation better than ~50-60%/$\sqrt{E}$ on hadrons, even with perfect scintillator/Cerenkov separation.

(8) *To achieve the theoretical/MC minimum energy resolution:* ~15%-18%/$\sqrt{E}$ on jets, separate scintillator sensors with $h/e_i$ ~ 0.6-0.8 (likely hydrogenous), and Cerenkov sensors with $h/e_C \leq 0.2$ are needed, which we will both explore in MC and test in this proposal. To achieve $h/e_C < 0.2$-0.25, lower n (index of refraction) Cerenkov radiators are required (i.e. $\beta_{thresh}$ ->1), but sufficiently thick for enough photons to achieve an e-m resolution <60%/$\sqrt{E(GeV)}$ or $N_{pe}$>3 pe/GeV.

<div align="center">*Precis of Parallel Fiber Dual Readout Calorimeter Problematic Issues:*</div>

In this section we discuss the issues with Parallel Cerenkov and Scintillator fiber Dual Readout. Parallel fiber calorimetry using plastic scintillator fibers and quartz fibers has intrinsic limitations beyond the small overlap of hadronic and e-m responses. We briefly summarize these deficits:

1. *Constant Term* – unavoidable issue – scintillator light attenuation in ~2m of scintillating fibers leads to uncorrectable longitudinal variations in the signal.

2. *Pointing/Projective Geometry* problematic in a practical parallel fiber calorimeter over a substantial solid angle. The mechanics + fiber packing of fully projective (θ,ϕ) very difficult for (pitch,yaw) more than a few degrees. Streaming down fiber holes lowered the resolution in DREAM, *even at a 2° pitch. (5° ~ η=3 at LHC).* Packing extra parallel fibers from the back or obtaining 2m conical fibers to obtain projective wedges with the same sampling fraction would still lead to a larger constant term and calibration difficulties.

3. *Scintillator Fiber & Photodetector Raddam:* At present, there are no good examples of fast decay scintillator fibers which have *proven* sufficient raddam resistance to be useful for hadron calorimetry at most future colliders or very high rate experiments (see the end remark below).

4. *Fiber Bundle & Photodetector Punchthrough:* Huge fiber bundles, >33% of the back of the fiber dual calorimeter area, are directly behind the calorimeter. Large punchthrough backgrounds are generated by these fibers, photodetectors (~1/800 incident π/K quasi-elastic scatter through a 10 Lint calorimeter). (see figures 2 below)

4. *E-M and Hadronic Components of Incident Jets:* Parallel fibers almost no ability to detect the separate incident direct e-m components of a jet, since there is no longitudinal segmentation.

5. *High Resolution EM Front End.* Parallel fiber dual readout jet calorimeter by itself has no ability to make a compensated high-Z high sampling EM front end.

6. *Calibration:* Parallel fiber geometry difficult to calibrate, as radiation damage & attenuation varies along the length of the fibers, in contrast with longitudinally segmented calorimeters.

7. *Timing & Pileup:* Longitudinal fibers store the information of jet/em showers: the signal is over the time for the light to traverse the fibers. The light generated at the back of the calorimeter arrives at the photodetector first. Fiber calorimeters measure the falling edge of the shower, a less precise measurement

8. *Longitudinal Segmentation:* Fiber dual readout is incompatible with true longitudinal segmentation even with waveform electronics, and cannot be easily rebuilt for front raddam or implement 4,5 above.

9. *Cerenkov (Fiber) Index of Refraction:* High Radiation Resistant Cerenkov fibers are limited at present to quartz, with n=1.46 – $h/e_C$ ~0.25-0.20 – limiting resolution. Low index n tiles with even lower $h/e_C$ ratio are possible and include n<1.1 silica aerogels, n~1.31 Teflon AF (the amorphous form of Teflon which is radiation reisstnat to at least 10 MRad), n~1.38 Siloxanes, n~1.4 Fluoride glasses, and others.

10. *Particle Flow/Energy Flow High Granularity Calorimetry:* improving jet δθ/θ, core ID of jets, isolation/ID of leptons/photons in jets and pileup, and neutral particle ($K^O$, n) ID, all especially under pileup. Tile multiple readouts are compatible with particle flow corrections, and can be easily added to particle flow calorimeters, whereas parallel fiber readouts are incompatible.

11. *Other Sensors for Dual & Multiple Readout:* Parallel fiber geometries cannot readily use other hadronic or n-enhanced sensors, and other e-m energy sensors. Examples of other detectors which are better in tile form factors and which have radiation resistance include: drifted ionization detectors (solid – Si, Diamond, GaAs; liquid- LArgon; gasses – micromegas..); β->1 sensitive detectors such TRD, or ultra-low-index materials(aerogels, $MgF_2$, water, perfluoro compounds, silicones, and others); secondary emission sensors with higher response to slow particles β-> 0 and minimal response to minimum ionizing energy (new large MCP); inorganic non-hydrogenous scintillators (LYSO, PbWO4 et al.), and $^6$Li, $^{10}$B or $^3$He containing materials for neutron signal enhancement.

12. *Cost:* the cost of tiles is significantly less per mass or volume of sensitive material than that of fibers, and the cost and complexity of a fabricated tile absorber matrix is considerably less than the parallel fiber Swiss cheese absorber.

13. **Homogenous Calorimeters** reading out both scintillation and Cerenkov light based on timing perforce must have less contrast between e-m and hadronic howers if based on timing differences between prompt Cerenkov and Scintillation as both give some signal at the same times, and if the scintillator is negligible at early times, the calorimeter will be too slow to work at 40-100 MHz. Similarly, measurements based on bandpass filters will have overlap. Additionally, such proposals are not amenable to particle/energy flow algorithms if using large homogeneous crystal scintillators.

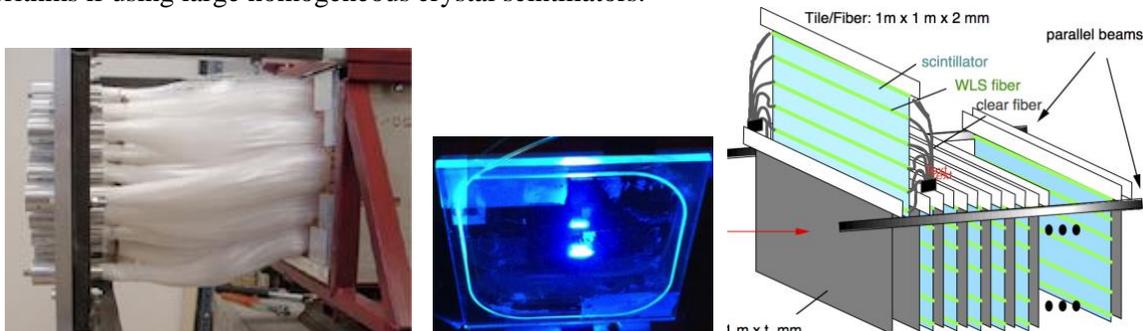

*Figures : from Left: Fiber bundles exiting the Dream Calorimeter to PMT – a source of noise and resolution loss;* **Sigma Tile***; Tile Calorimeter cartoon (hanging file, for testing multiple tile types).*

*- Radiation Damage:*

One "brute force" strategy for radiation resistance is developing easily replaceable parts/elements, especially with the remarkable advances in robotics for much faster and more dexterous than human replacement, even in beam conditions, and without human radiation exposure. Such replacement is not very feasible with parallel fiber calorimeters, but more amenable to tile sensor systems.

Calorimeters with as cassettes or cartridges of SiPM and/or tiles, fluids & gasses, and other replacement schemes are feasible if such backstops become ultimately necessary for future collider experiments. The development of potentially raddam-resistant large bandgap SiPM-like (APD pixel arrays) but based on other III-V or II-VI semiconductors that in principle could be embedded in calorimeters, and novel rad-resistant scintillator tiles (ZnO:Ga, Silicone-based, new transparent colorless polyimides/kaptons, quartz tiles melt-infused with inorganic nano-scintillators, and others) are not considered here but should be studied as future possibilities.

*Tile Dual MC:*

To have an idea if we can address the above issues with tiles, we performed a GEANT4 MC on a simple Dual Readout tile calorimeter, similar to those that could be used in large experiments to confirm that at minimum dual readout could be considered. A calorimeter consisting of 5mm thick each of quartz tiles, plastic scintillator tiles, and Cu absorber tiles was modeled. Two energies (50, 100 GeV) each of 1000 electrons (red dots) and of ~1000 pions (blue dots) [Figure 2a] were sent into the 50x50 cm area calorimeter, 12.2 $L_{int}$ (3.5 m) deep. The number of photons between 325-650nm generated in the Cerenkov tiles and in the scintillator (PPO-POPOP spectra) tiles were counted, and, in this toy model, 0.5% of the photons at random were assumed to be able to be collected and converted to p.e., consistent with present tile calorimeters with "sigma" tiles and WLS fiber readout. The collected scintillator photons were about 120x the Cerenkov tile photons, but photostatistics are not the most limiting factors. The means of the histograms of the electron shower p.e. in quartz and in scintillator were used to convert/normalize the number of collected p.e. in Cerenkov light and in Scintillator light to the same energies $E_{Cerenkov}$ and $E_{Scintillator}$, and then plotted as a scatter plot of $E_C$ vs $E_S$ for each electron[Figure 2a]. After that normalizing of the energies, the electron-incident scatter plot $E_c$ vs $E_s$ (red points) lie along the green line shown as $E_c = E_s$. The tight clustering is evidence of good energy resolution for both electromagnetic signals, but with the scintillator resolution clearly better (narrower). Pions of 50, 100 GeV were then simulated. The resulting $E_c$ vs $E_s$ signals were normalized to energy using the electron normalization and scatter-plotted. The hadron points(blue) in the scatter plot lie mainly below the $E_c = E_s$ electron line(in blue), with a clear correlation between $E_c$ vs $E_s$.

A line was fitted to the correlated hadron scatter points (Green line as shown schematically only for 50 GeV), and the fitted slope R of that line is used to correct the energy. The angle θ between the line $E_C=E_S$ and the green line linear fit with slope R to the blue hadron scatter points is given by θ= arctan(R)–π/4, shown as an arc (Fig2a). If one projects the blue scatter points as a histogram onto an energy axis perpendicular to the fitted green linear correlation line, the energy distribution becomes quite Gaussian and narrower. The true energy E is given *to first order* by $E_s$, plus a correction term proportional to the difference ($E_s-E_c$) as $E = E_s + \alpha(E_s-E_c)$ where α is given by the fitted slope R: R=(1+ α)/α or α =1/(1-R)[12] As the slope R gets steeper/larger, the correction linear term α($E_s-E_c$) becomes more important as $E_C$ falls faster than $E_s$. What this correction says is that when the Cerenkov energy $E_c$ is the same as scintillation energy $E_s$, as is the average case with electrons or $\pi^{+/-}$ charge-exchange to $\pi^o$, then ($E_s-E_c$)~> 0 and no correction is needed to $E_s$, the 1st approximation to the energy. The difference ($E_s-E_c$) grows as the shower fluctuates more into nuclear/hadronic energies, and $E_s$ must be increased by an amount proportional to ($E_s-E_c$), with a proportionality constant α =1/(1-R), in effect linearly projecting the scatter plot onto a histogram with an axis perpendicular to the fitted line. *The (mean, rms) = (100, 2.66) GeV [Fig. 2b] shows energy resolution promise that could enable W-> jet-jet separation as in Fig 1b, especially with even higher sampling frequency (1/5-1/10 Xo) that will be performed in this proposal.*

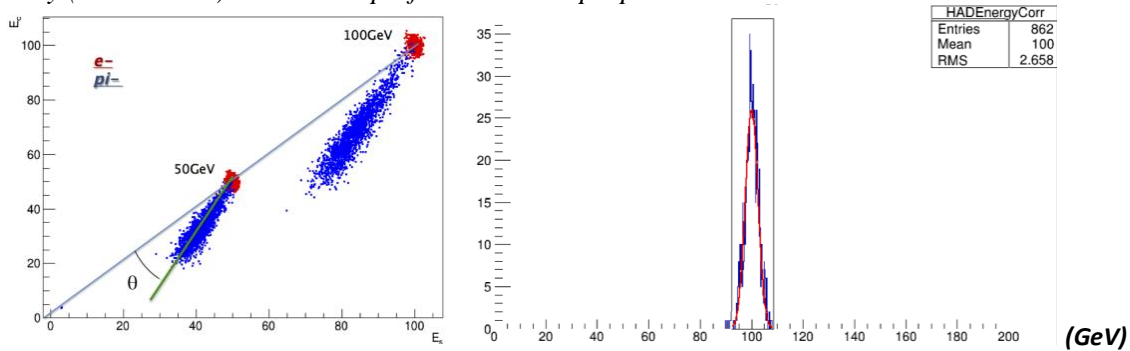

**Fig. 2a (L):** Tile Dual Readout GEANT4 MC: Scatter plot of $E_{Cerenkov}$ ($E_C$) vs $E_{Scintillator}$ ($E_s$) in a tile calorimeter consisting of tiles 0.5 cm thick each of quartz, plastic scintillator, and Cu absorber. Two

energies (50, 100 GeV) each of electrons (red dot clusters on line $E_c=E_s$) and pions (blue dots) with a linear fit (green line shown). When the pion ($E_s,E_c$) points were projected onto an axis perpendicular to the linear fit, the resolution on hadrons is narrower and more Gaussian, shown for 100 GeV $\pi$ in Fig2b.

**Fig. 2b(R):** Histogram of the 100 GeV $\pi$ events at left vs Cerenkov corrected energy, in effect projected perpendicularly to the linear fit to the $E_c$ vs $E_s$ correlation. *The (mean, rms) = (100, 2.66) GeV shows promise for W-> jet-jet separation (see Fig 1b).* Finer sampling (0.2 $X_o$) will improve this resolution.

It will be possible enhance this simple linear fit to the scatter plot for dual readout correction to include curvature with higher order fitted terms $\alpha_2(E_s-E_c)^2 + \alpha_3(E_s-E_c)^3 +..$, with energy dependent $\alpha$'s – there is a continuous mapping of the correlations in $E_c$ vs $E_s$ space to the line $E_c=E_s$.

The preceding results and information lead us to propose studying multiple readouts with tiles. We want to know if the following ideas improve calorimeter resolution:

*Multiple Tile Readout – Beyond Dual Tile - Triple Readout and beyond*
- *Adding sensor tiles* with characteristics and higher contrast between scintillation and Cerenkov - different than those of quartz and plastic scintillator improves resolution – tiles relatively more insensitive to MIPs, OR more sensitive to $\gamma\beta$->0 increases the contrast between e-m and hadronic energy (enhancing the low energy hadronic signal) - *Multiple Readout.*

- Extending Dual Readout to include 3 or more sensors to see if multidimensional correlations of sensors can yield an algorithm to improve energy resolution beyond dual readout. We propose non-hydrogenous scintillator, hydrogenous or other neutron-sensitive scintillator, and 2 indices of Cerenkov tile(s), or the SE tiles above, that may enable even more correction, by also comparing less-sensitive neutron scintillators (direct charged particles including nuclear fragments) such as non-hydrogenous scintillators (inorganic and perfluorcompounds) to more neutron-sensitive scintillator tiles. Included last in the systematic MC studies will be combining the best properties of both tile dual readout and particle flow, by developing collaborations with Energy/Particle Flow(PF)[13,14] R&D groups to add $\beta$−sensitive tiles Cerenkov tiles to PF tile prototypes.

- *Theoretical Multiple Readout Resolution Targets:* Early MC and Back-of-the envelope extrapolations indicate that ~15%-18%/$\sqrt{E}$ is possible on jets: scintillator sensors with $h_i/e_i$ ~ 0.6-0.8 (lhydrogenous & n-sensitive), and Cerenkov sensors with $h_c/e_C \leq 0.2$ are needed. To achieve $h_c/e_C < 0.2$, lower index of refraction Cerenkov radiators are required (i.e. $\beta_{thresh}$ ->1), but require enough thickness for photons to achieve an e-m resolution < 70%/$\sqrt{E(GeV)}$ or Npe > 2 pe/GeV, and careful adjusting of both $f_{sam}$ and $f_{freq}$.

*Multiple Readout Tiles:*
We propose a systematic investigation of the many possible variations of Cerenkov tile dual readout, with
- 1. alternatives to Cerenkov light as the compensation 2$^{nd}$ piece of shower information not possible with practical existing fibers, and
- 2. extended to Multiple Tile Readout, not only via plastic scintillator and transparent Cerenkov tiles, with varied packing fractions and sampling, but a large variety of other types of sensor tiles with different responses to neutrons, ion fragments, and 0<$\beta$<1 sensitive tiles.

These sensors include more radhard, hydrogenous, non-hydrogenous, neutron-sensitive-enhanced tiles, ion detector "tiles", ion-collecting liquids, TRD tiles, Secondary Emission, and others. Including a 3$^{rd}$ or 4$^{th}$ sensor plane with different neutron (hydrogenous vs non-hydrogenous) or $\beta$->0 sensitivity may enable more corrections for lost nuclear energy. Rate and timing capability is a serious issue –at SLHC and beyond, FW10%-10%Max in less than ~10ns with 200 pileup is advantageous; where appropriate, we will include in the MC studies sensors with high rate, timing & pileup capability, but concentrate in this Task on understanding and extending tile dual readout and Particle Flow. MC variations that may be considered include the following:

*a) Fused silica tiles* (n=1.46; h/e$_C$ ratio ~0.25) in a hadron calorimeter with very fine sampling (0.3-3mm)

*b) Lower index tiles (*1.05<n<1.35): rad-hard tiles to achieve a high contrast ratio with h/e$_C$->0.15: silica aerogels (n=1.05-1.3), polysiloxanes (n=1.35, 100 MRad), TeflonAF (n=1.29, the water-clear amorphous form of Teflon with at least 12 MRad hardness), MgF$_2$(n=1.37).

*c) TRD* (Transition Radiation Detector, such as straw tubes with a high β threshold) tiles for detecting the e-m component, interspersed with scintillator tile and absorbers;

*d) Bulk Liquids:* for very large detectors such as long-baseline or cosmic neutrinos.
- d.1 LArgon drifted ions and simultaneous Cerenkov light detection. The index is low enough that a good e/h contrast seems possible, and the scintillation light at 128nm will not penetrate light detector windows.

- d.2 water "tiles" using n=1.29-1.31 TeflonAF films for light piping + liquid scintillator "tiles";

*e) Drifted Ion detectors* – "tiles" of Si, LArgon, high pressure hydrogenous gas mixtures (methane, butane).

*f) Secondary Emission (SE).* Large glass-based thin-film activated MCP[15] or metal foil "dynodes" or metal-oxide films that are directly sensitive to ionizing particles, and could serve as a "tile". The metal etched dynodes used in Hmamatsu PMT are made in large sheets which are diced. SE is fast (10-20 ps to generate SE electrons) and exceptionally radiation resistant. The beam monitors at LHC take $10^{20}$ MIP protons-cm$^{-2}$ without any noticeable degradation, and similarly the metal oxide dynodes in PMT have effectively gigarad of exposure. Metal oxide films in general are exceptionally radiation resistant – such as Alumina. Secondary Emission (SE) tiles are more sensitive to γβ-›0 particles than to MIPs - the SE signal scales as dE/dx (βγ), with a MIP SE signal~100-200x less than that of the particle at the energy of the peak SE signal – *the opposite of Cerenkov light*[16] (see Fig 3 below). SE tiles may provide further correction for binding energy effects and lost neutron energies. Figs. 3 shows secondary yield vs particle energy – the yield peaking at low energy is exceptional for ion fragments and neutron knock-ons.

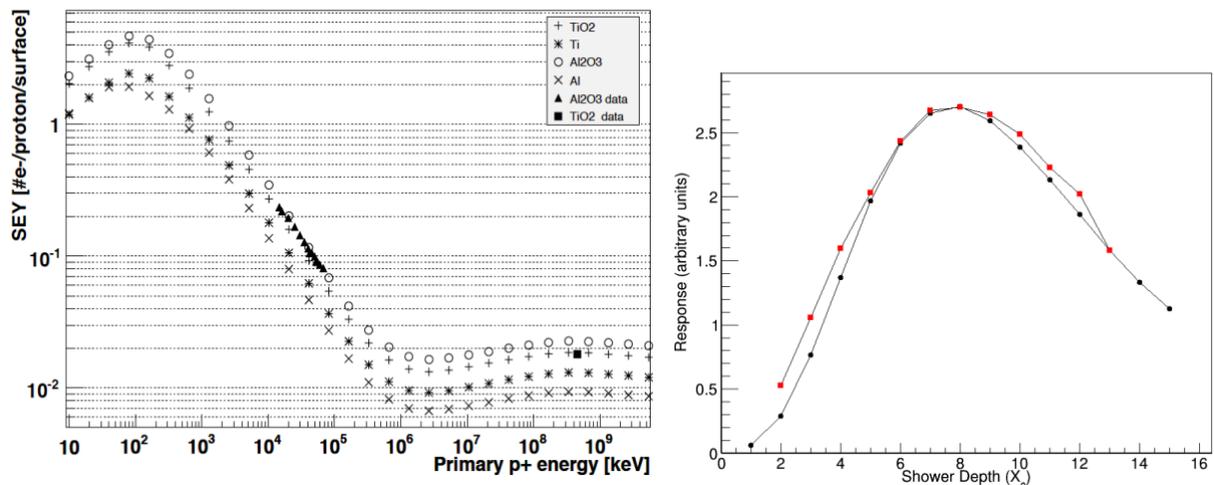

*Figures 3a,b:* (L) The *secondary yield* of protons on Al$_2$O$_3$, TiO$_3$, Al, Ti films vs proton energy *in KeV* (data: LHC beam monitors). The peak yield is at ~100-200 KeV falls by x~200 at minimum (~2 GeV), *scaling as βγ* (electron peak ~150 eV, ~PMT dynodes) *the opposite of Cerenkov light*. Low energy sensitive/high energy suppressed response may improve multiple tile readout compensation. '

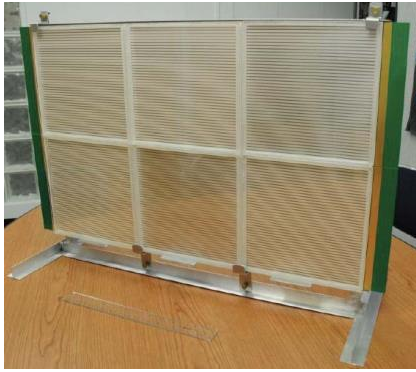

*Fig 3c:* 24"x16" MCP at ANL/Chicago in charged particle stimulation of the MCP at the Fermilab Test Beam has responses completely consistent with the SE yield shown above.

**g) E-M Compartment**: Homogeneous, high index dense scintillator e-m front sections reduce jet energy resolution substantially, but cannot be fully compensated by their own Cerenkov light. We propose study in MC Shashlik[17,18]-like Calorimeters consisting of non-hydrogenous high light output scintillating tiles (LYSO, $PbWO_4$, ..), sub-Xo thick W tiles, interspersed with low index Cerenkov tiles with $h/e_C$ ratio < 0.15, such as $MgF_2$, Teflon AF(10 MRad rad resistance), silica aerogels n=1.05-1.1[19]. (WLS readout fibers would be periodically blinded to the wrong tiles).

**h) *Triple Readout and beyond:*** Extending Dual Readout to include 3 or more sensors to see if multidimensional correlations of sensors can yield an algorithm to improve energy resolution beyond dual readout. We propose non-hydrogenous scintillator, hydrogenous or other neutron-sensitive scintillator, and 2 indices of Cerenkov tile(s), or the SE tiles above, that may enable even more correction, by also comparing less-sensitive neutron scintillators (direct charged particles including nuclear fragments) such as non-hydrogenous scintillators (inorganic and perfluorcompounds) to more neutron-sensitive scintillator tiles. Included last in the systematic MC studies will be combining the best properties of both tile dual readout and particle flow, by developing collaborations with Energy/Particle Flow(PF)[20,21] R&D groups to add $\beta$−sensitive tiles Cerenkov tiles to PF tile prototypes.

**Future Directions:**
- Tiles may be optimized for multiple readout to increase neutron detection (incorporating n-detecting isotopes $^{10}B$, $^6Li$, $^6LiH$, $^6Li^{10}BH_4$, $^{235}U$, and $^{235}U^{10}B_2$ in scintillator or semiconductor sensors), and ion fragment detection (thin detectors <30μm) on the surface of absorber tiles, largely insensitive to MIP particles.

*Summary:*
We propose extending parallel fiber dual readout calorimetry to tiles, more applicable to many future experimental requirements, with superior energy resolution. Monte Carlo (MC) studies indicate that a tile dual calorimeter including an integral Cerenkov-compensated e-m front end and further longitudinal segmentation, not possible with parallel fibers, has equivalent or better resolution. Besides comparison and tuning with dual tile calorimeter data, a MC can be extended to study other dual and then multiple tile sensors including tiles with higher contrast to em-hadron shower fluctuations with low refractive indices (much lower than quartz or plastic), transition radiation, secondary emission, hydrogenous/non-hydrogenous ionization-sensing, and neutron and ion-fragment sensing tiles for improving dual readout not available with fibers, and beyond to triple or more readout. For example, secondary emission tiles (like dynodes) are very sensitive to ion fragments and low energy neutrons. We suggest MC studies for adding Cerenkov and other tiles to Particle Flow/High Granularity tile calorimeters such CALICE and planned in CMS & ATLAS Phase III upgrades, groups studying future machine(ee,pp,ep) detectors, b-physics, tagged neutrino experiments, and space-based calorimeters.